\documentclass[journal]{IEEEtran}
\usepackage{color}
\usepackage{times}

\usepackage{soul}
\usepackage{url}
\usepackage[hidelinks]{hyperref}
\usepackage[utf8]{inputenc}
\usepackage[small]{caption}
\usepackage{graphicx}
\usepackage{amsmath}
\usepackage{booktabs}
\usepackage{amsfonts}
\usepackage{subcaption} 

\ifCLASSINFOpdf
\else
\fi
\begin{document}
%
\title{CycleDRUMS: Automatic Drum Arrangement For Bass Lines Using CycleGAN}
%
%
%

\author{Giorgio~Barnabò,
        Giovanni~Trappolini,
        Lorenzo~Lastilla,
        Cesare~Campagnano,
        Angela~Fan,
        Fabio~Petroni,
        and~Fabrizio~Silvestri}

\markboth{IEEE TRANSACTIONS ON MULTIMEDIA}%
{Shell \MakeLowercase{\textit{et al.}}: Bare Demo of IEEEtran.cls for IEEE Journals}
%



\maketitle

\begin{abstract}
The two main research threads in computer-based music generation are: the construction of \textit{autonomous music-making systems}, and the design of \textit{computer-based environments to assist musicians}. In the symbolic domain, the key problem of automatically arranging a piece music was extensively studied, while relatively fewer systems tackled this challenge in the audio domain. In this contribution, we propose CycleDRUMS, a novel method for generating drums given a bass line. After converting the waveform of the bass into a  mel-spectrogram, we are able to automatically generate original drums that follow the beat, sound credible and can be directly mixed with the input bass. We formulated this task as an \textit{unpaired image-to-image translation} problem, and we addressed it with CycleGAN, a well-established unsupervised style transfer framework, originally designed for treating images.
The choice to deploy raw audio and mel-spectrograms enabled us to better represent how humans perceive music, and to potentially draw sounds for new arrangements from the vast collection of music recordings accumulated in the last century.
In absence of an objective way of evaluating the output of both generative adversarial networks and music generative systems, we further defined a possible metric for the proposed task, partially based on human (and expert) judgement. 
Finally, as a comparison, we replicated our results with Pix2Pix, a paired image-to-image translation network, and we showed that our approach outperforms it.

\end{abstract}


%
\IEEEpeerreviewmaketitle

\section{Introduction}
%
%
%
%

\IEEEPARstart{T}{he} development of home music production has brought significant innovations into the process of pop music composition. Software like Pro Tools, Cubase, and Logic -- as well as MIDI-based technologies and digital instruments -- provide a wide set of tools to manipulate recordings and simplify the composition process for artists and producers.
After recording a melody, maybe with the aid of a guitar or a piano, song writers can now start building up the arrangement one piece at a time, sometimes not even needing professional musicians or proper music training. As a result, singers and song writers -- as well as producers -- have started asking for tools that could facilitate, or to some extent even automate, the creation of full songs around their lyrics and melodies. To meet this new demand, the goal of designing computer-based environments to assist human musicians has become central in the field of automatic music generation \cite{briot2020deep}. IRCAM, \cite{assayag1999computer}, 
Sony CSL-Paris FlowComposer, \cite{papadopoulos2016assisted}, and Logic Pro X Easy Drummer are just some examples. In addition, more solutions based on deep learning techniques, such as RL-Duet \cite{jiang2020rl} -- a deep reinforcement learning algorithm for online accompaniment generation -- or PopMAG, a transformer-based architecture which relies on a multi-track MIDI representation of music \cite{ren2020popmag}, continue to be studied. A comprehensive review of the most relevant deep learning techniques applied to music is provided by \cite{briot2020deep}. 

Unlike most techniques that rely on a symbolic representation of music (i.e. MIDI, piano rolls, music sheets), the approach proposed in this paper is a first attempt at automatically generating drums in the audio domain, given a bass line encoded in the mel-spectrogram time-frequency domain. Needless to say, as extensively shown in section \ref{related_works}, mel-spectrograms are already commonly and effectively used in many music information retrieval tasks \cite{4895319}. Nonetheless, music generation models applied to this kind of intermediate representation are still relatively scarce. Although arrangement generation has been extensively studied in the context of symbolic audio, switching to mel-spectrograms allowed us to preserve the sound heritage of other musical pieces and certainly represents a valid alternative for real-case scenarios. Indeed, even if it is possible to use synthesizers to produce sounds from symbolic music, MIDI, music sheets and piano rolls are not always easy to find or produce, and they sometimes lack in expressiveness. Moreover, even state-of-the-art synthesizers cannot yet reproduce the infinite nuances of real voices and instruments, whereas raw audio representation guarantees more flexibility and requires little music competence.
On the other hand, thanks to this two dimensional time-frequency representation of music based on mel-spectrograms, we can treat the problem of automatically generating an arrangement or accompaniment for a specific musical sample equivalent as an image-to-image translation task. For instance, if we have the mel-spectrogram of a bass line, we may want to produce the mel-spectrogram of the same bass line together with suitable drums. 

To solve this task, we tested an unpaired image-to-image translation strategy known as CycleGAN \cite{zhu2017unpaired}. 
In particular, we trained a CycleGAN architecture on 5s bass and drum samples (equivalent to $256\times256$ mel-spectrograms) coming from both the Free Music Archive (FMA) dataset \cite{fma_challenge}, and the musdb18 dataset \cite{musdb18}. The short sample duration does not affect the proposed methodology, at least with respect to the arrangement task we focus on, and inference could be performed on longer sequences as well. Since the FMA songs lack source separated channels (i.e. differentiated vocals, bass, drums, etc.) it was pre-processed first. The required channels were extracted using Demucs \cite{defossez2019demucs}. The results obtained were then compared to Pix2Pix \cite{isola2017image}, another popular paired image-to-image translation network. To sum up, our main contributions are the following ones: 
\begin{itemize}
\item we trained a CycleGAN architecture on bass and drum mel-spectrograms in order to automatically generate drums that follow the beat and sound credible for any given bass line;
\item our approach is able to generate drum arrangements with low computational resources and limited inference time, if compared to other popular solutions for automatic music generation \cite{dhariwal2020jukebox};
\item we developed a metric -- partially based on or correlated to human (and expert) judgement -- to automatically evaluate the obtained results and the creativity of the proposed system, given the challenges of a quantitative assessment of music;
\item we compared our method to Pix2Pix, another popular image transfer network, showing that the music arrangement problem can be better tackled with an unpaired approach and adding a cycle-consistency loss.
\end{itemize}
To the best of our knowledge, we are the first to exploit cycle-consistent adversarial networks and a two dimensional time-frequency representation of music for automatically generating suitable drums given a bass line. 

\section{Related Works}
\label{related_works}

The interest surrounding automatic music generation, translation and arrangement has greatly increased in the last few years, as proven by the high number of solutions proposed -- see \cite{briot2020deep} for a comprehensive and detailed survey. Here we present a brief overview of the key contributions both in symbolic and audio domain. 

\textbf{Music generation \& arrangement in the symbolic domain.} There is a very large body of research that uses symbolic music representation to perform music generation and arrangement. The following contributions used MIDI, piano rolls, chord and note names to feed several deep learning architectures and tackle different aspects of the music generation problem. 
In \cite{yang2017midinet}, CNNs are used for generating melody as a series of MIDI notes either from scratch, by following a chord sequence, or by conditioning on the melody of previous bars, whereas in \cite{mogren2016c,mangal2019lstm,jaques2016tuning,makris2017combining} LSTMs are used to generate musical notes, melodies, polyphonic music pieces, and long drum sequences under constraints imposed by metrical rhythm information and a given bass sequence. The authors of \cite{yamshchikov2017music,roberts2018hierarchical, lattner2019highlevel} instead, use a variational recurrent auto-encoder to generate melodies. In \cite{boulanger2012modeling}, symbolic sequences of polyphonic music are modeled in a completely general piano-roll representation, while the authors of \cite{hadjeres2017interactive} propose a novel architecture to generate melodies satisfying positional constraints in the style of the soprano parts of the J.S. Bach chorale harmonisations encoded in MIDI. In \cite{johnson2017generating}, RNNs are used for the prediction and composition of polyphonic music; in \cite{hadjeres2017deepbach}, highly convincing chorales in the style of Bach were automatically generated using note names; \cite{lattner2018imposing} added higher-level structure on generated polyphonic music, whereas in \cite{mao2018deepj} an end-to-end generative model capable of composing music conditioned on a specific mixture of composer styles  was designed. The approach described in \cite{hawthorne2018enabling}, instead, relies on notes as an intermediate representation to a suite of models -- namely, a transcription model based on a CNN and a RNN network\cite{hawthorne2017onsets}, a self-attention-based music language model \cite{huang2018music} and a WaveNet model \cite{oord2016wavenet} -- capable of transcribing, composing, and synthesizing audio waveforms. Finally, \cite{zhu2018xiaoice} proposes an end-to-end melody and arrangement generation framework, called XiaoIce Band, which generates a melody track with multiple accompaniments played by several types of instruments. 

\textbf{Music generation \& arrangement in the audio domain.} Some of the most relevant approaches proposed so far in the field of waveform music generation deal with raw audio representation in the time domain. Many of these approaches draw methods and ideas from the extensive literature on audio and speech synthesis.  
For instance, in \cite{prenger2019waveglow} a flow-based network capable of generating high quality speech from mel-spectrograms is proposed, while in \cite{wang2019neural} the authors present a neural source-filter (NSF) waveform modeling framework that is straightforward to train and fast to generate waveforms. In \cite{zhao2020transferring} recent neural waveform synthesizers such as WaveNet, WaveG-low, and a neural-source-filter (NSF) are compared. \cite{mehri2016samplernn} tested a model for unconditional audio synthesis based on generating one audio sample at a time, and \cite{bhave2019music} applied Restricted Boltzmann Machine and LSTM architectures to raw audio files in the frequency domain in order to generate music, whereas the authors of \cite{oord2016wavenet} propose a fully probabilistic and auto-regressive model, with the predictive distribution for each audio sample conditioned on all previous ones, to produce novel and often highly realistic musical fragments. The authors of \cite{manzelli2018end} present a raw audio music generation model based on the WaveNet architecture, which takes as a secondary input the notes of the composition.
Finally, in \cite{dhariwal2020jukebox} the authors tackled the long context of raw audio using a multi-scale VQ-VAE to compress it to discrete codes, and modeled such context through Sparse Transformers, in order to generate music with singing in the raw audio domain. Nonetheless, due to the computational resources required to directly model long-range dependencies in the time domain, either short samples of music can be generated or complex and large architectures and long inference time are required.
On the other hand, in \cite{vasquez2019melnet}, authors discuss a novel approach which proves that long-range dependencies can be more tractably modelled in two-dimensional time-frequency representations such as mel-spectrograms. More precisely, the authors of this contribution designed a highly expressive probabilistic model and a multi-scale generation procedure over mel-spectrograms capable of generating high-fidelity audio samples which capture structure at timescales.  It is worth recalling, as well, that treating spectrograms as images is the current standard for many Music Information Retrieval tasks, such as music transcription \cite{sigtia2016end}, music emotion recognition \cite{dong2019bidirectional} and chord recognition.

\textbf{Generative adversarial networks for music generation.} Such two-dimensional representation of music paves the way to the application of several image processing techniques and image-to-image translation networks to carry out style transfer and arrangement generation \cite{isola2017image,zhu2017unpaired}. It is worth recalling that the application of GANs to music generation tasks is not new: in \cite{brunner2018symbolic}, GANs are applied on symbolic music to perform music genre transfer, while in \cite{7254179, nistal2020drumgan}, authors construct and deploy an adversary of deep learning systems applied to music content analysis; however, to the best of our knowledge, GANs have never been applied to raw audio in the mel-frequency domain for music generation purposes. As to the arrangement generation task, the large majority of approaches proposed in the literature is based on a symbolic representation of music: in \cite{ren2020popmag}, a novel multi-track MIDI representation (MuMIDI) is presented, which enables simultaneous multi-track generation in a single sequence and explicitly models the dependency of the notes from different tracks by means of a Transformer-based architecture; in \cite{jiang2020rl}, a deep reinforcement learning algorithm for online accompaniment generation is described. 

Coming to the most relevant issues in the development of music generation systems, both the training and evaluation of such systems have proven challenging, mainly because of the following reasons: (i) the available datasets for music generation tasks are challenging due to their inherent high-entropy \cite{dieleman2018challenge}, and (ii) the definition of an objective metric and loss is a common problem to generative models such as GANs: as of now, generative models in the music domain are evaluated based on the subjective response of a pool of listeners, because an objective metric for the raw audio representation has never been proposed so far. Just for the MIDI representation a set of simple musically informed objective metrics was proposed \cite{yang2020evaluation}.

\section{Method}

We present CycleDRUMS, a novel approach for automatically adding credible drums to bass lines, based on an adversarially trained deep learning model.

\subsection{Source Separation for Music}

A key challenge to our approach is the scarce availability of music data featuring source separated channels (i.e. differentiated vocals, bass, drums, ...). To this end, we leverage Demucs by \cite{defossez2019demucs}, a freely available tool which separates music into its generating sources. Demucs is an extension to Conv-Tasnet \cite{Luo2019ConvTasNetSI}, purposely adapted to the field of music source separation. It features a U-NET encoder-decoder architecture with a bidirectional LSTM as 
hidden layer. In particular, we exploited the authors' pre-trained model consisting of 6 convolutional encoder and decoder blocks and a 
hidden size of length $3200$. Thanks to the randomized equivariant stabilization, Demucs is time-equivariant, meaning that any shifts in the input mixture will cause congruent shifts in the output. 

A potential weakness of this method, however, is that it sometimes produces noisy separations, with watered-down harmonics and traces of other instruments in the vocal segment. If follows that the usage of Demucs could somehow hinder our pipeline from properly recognising and  reconstructing the accompaniment, where the harmonics play a critical part. Nonetheless, even if better source-separation methods are available, achieving slightly higher values of signal to distortion ratio (SOTA SDR = 5.85, Demucs SDR = 5.67), we chose to use Demucs because it is faster and easier to embed in our pipeline. Moreover, Demucs outperforms the current state of the art for bass source separation [SOTA SDR = 5.28, Demucs SDR = 6.21]. 

Thanks to Demucs, we were at least partially able to solve the challenge of data availability and to feed our model with appropriate signals. In practice, given an input song, we use Demucs to separate it into vocals, bass, drums, and other, keeping of course the original mixture.




\subsection{Music Representation -- from Raw Audio to Mel-spectrograms}

A distinguishing feature of our method is to use mel-spectrograms, instead of waveforms. Namely, we opted for a two-dimensional time-frequency representation of music, rather than a time representation. The spectrum is a common transformed representation for audio, obtained via a Short-Time Fourier transform (STFT) \cite{muller2015fundamentals}. The discrete STFT of a given signal $x:[0:L-1]:=\{0,1,\ldots,L-1\}\to{\mathbb R}$ leads to the $k^{\mathrm{th}}$ complex Fourier coefficient for the $m^{\mathrm{th}}$ time frame:

$$
\mathcal{X}(m,k) := \sum_{n=0}^{N-1} x(n+mH)\cdot w(n)\cdot e^{-\frac{2\pi ikn}{N}}
$$

With $m\in[0:M]$ and $K\in[0:K]$, and where $w(n)$ is a sampled window function of length $N\in\mathbb{N}$, and $H\in\mathbb{N}$ is the hop size that determines the step size the window is to be shifted across the signal \cite{muller2015fundamentals}. The spectrogram is a two-dimensional representation of the squared magnitude of the STFT, i.e. $\mathcal{Y}(m,k) := | \mathcal{X}(m,k)|^2$, with $m\in[0:M]$
and $K\in[0:K]$. 

\begin{figure*}
\begin{center}
\includegraphics[width=.8\textwidth]{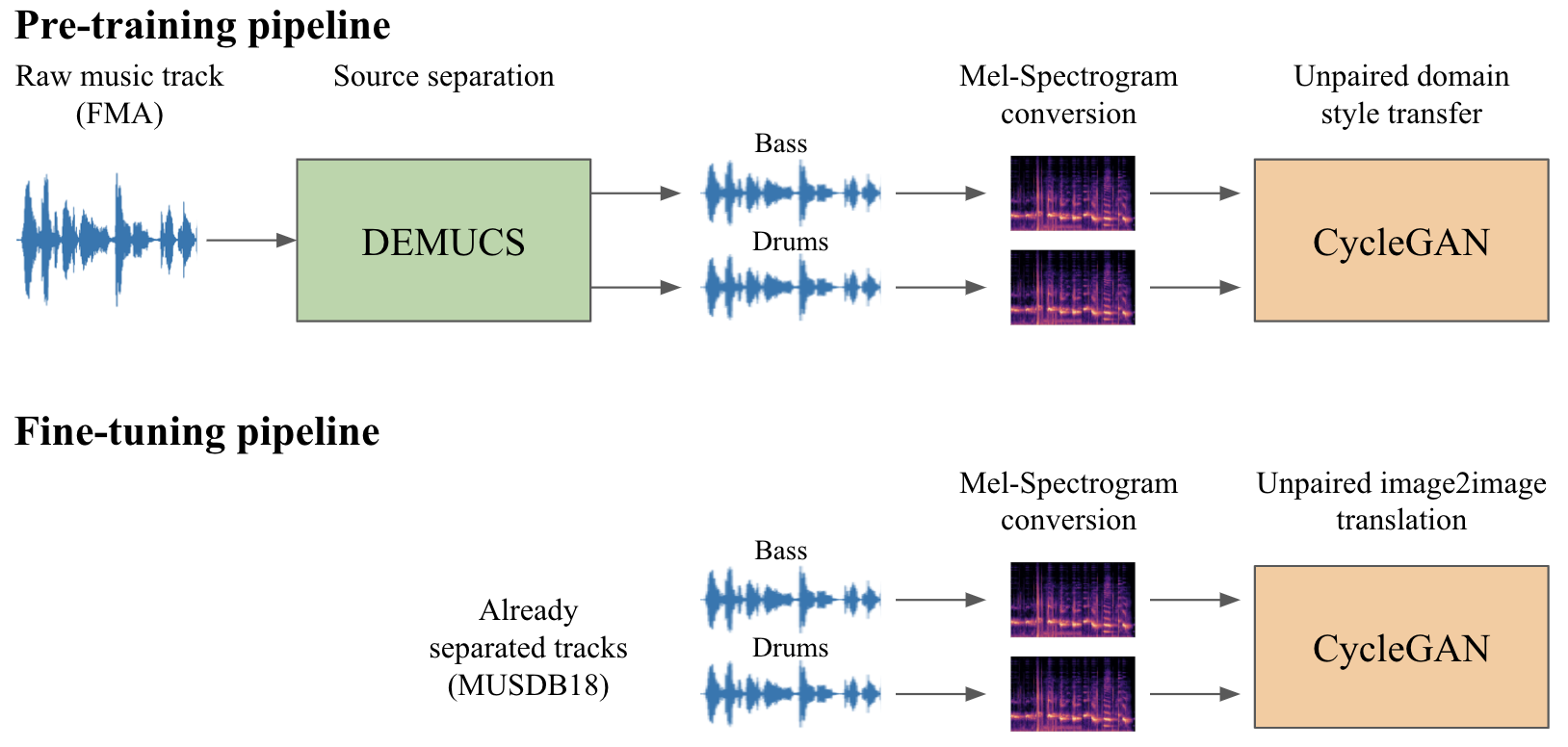}
\end{center}
\caption{To solve automatic drum arrangement task, we tested an unpaired image-to-image translation strategy known as CycleGAN \cite{zhu2017unpaired}. In particular, we trained a CycleGAN architecture on 5s bass and drum samples (equivalent to $256\times256$ mel-spectrograms) coming from both the Free Music Archive (FMA) dataset \cite{fma_challenge}, and the musdb18 dataset \cite{musdb18}. Since the FMA songs lack source separated channels (i.e. differentiated vocals, bass, drums, etc.), the bass and drum channels were extracted using Demucs \cite{defossez2019demucs}. After the source separation task is carried out on our song dataset, both the bass and drum waveforms are turned into the corresponding mel-spectrograms. The sampling rate was set to 22050 Hz, the window length $N$ to 2048, the number of mel-frequency bins to 256 and the hop size $H$ to 512. To fit our model requirements, we cropped out $256\times256$ windows from each mel-spectrogram with an overlapping of 50 time frames, obtaining multiple samples from each song (each roughly equivalent to 5 seconds of music). Since FMA is much larger than musdb18, but also lower-quality due to the artificial separation of sources, we used FMA to train the model, and then we fine-tuned it with musdb18 which comes in a source-separated fashion.}
\label{fig:Mel_SP}
\end{figure*}

Figure \ref{fig:Mel_SP} shows an example of a mel-spectrogram \cite{mel_scale} that is treated as a single channel image, representing the sound intensity with respect to time -- x axis -- and frequency -- y axis \cite{briot2020deep}. This decision allows to better deal with long-range dependencies, typical of such kind of data, and to reduce the computational resources and inference time required. Moreover, the mel-scale is based on a mapping between the actual frequency $f$ and perceived pitch $m = 2595 \cdot log_{10}(1 + \frac{f}{700})$, as the human auditory system does not perceive pitch in a linear manner. Finally, using mel-spectrograms of pre-existing songs to train our model potentially enables to draw sounds for new arrangements from the vast collection of music recordings accumulated in the last century. It is worth recalling that mel-frequency cepstral coefficients are the dominant features used in speech recognition, as well as in several music modeling tasks \cite{logan2001adaptive}.


After the source separation task is carried out on our song dataset, both the bass and drum waveforms are turned into the corresponding mel-spectrograms using PyTorch Audio\footnote{Available at: \url{https://pytorch.org/audio/stable/index.html}}. PyTorch works very fast and is optimized to perform robust GPU accelerated conversion. In addition, to reduce the dimensionality of the data, we decided to keep only the magnitude coefficients, discarding the phase information. Finally, in order to revert back the generated mel-spectrograms to the corresponding time-domain signal: (i.) we apply a conversion matrix (using triangular filter banks) that converts the mel-frequency STFT to a linear scale STFT. The matrix is calculated using a gradient-based method \cite{gradient_mel} that minimizes the Euclidean norm between the original mel-spectrogram and the product between reconstructed spectrogram and filter banks; (ii.) we use the Griffin-Lim's algorithm \cite{griffin_lim} to reconstruct the phase information.

It is worth noticing that the mel-scale conversion and the removal of STFT phases respectively discard frequency and temporal information, thus resulting in a distortion in the recovered signal. To minimize this problem, we made use of high-resolution mel-spectrograms \cite{vasquez2019melnet}, whose size can be tweaked with \textit{number of mels} and \textit{STFT hop size} parameters. Thus, here are the hyper-parameters we used: the sampling rate was set to 22050 Hz, the window length $N$ to 2048, the number of Mel-frequency bins to 256 and the hop size $H$ to 512. To fit our model requirements, we cropped out $256\times256$ windows from each mel-spectrogram with an overlapping of 50 time frames, obtaining multiple samples from each song (each roughly equivalent to 5 seconds of music). 

\subsection{Image to Image Translation - CycleGAN}
We casted the automatic drum arrangement generation task as an \textit{unpaired image-to-image translation} task, and we then solved it by adapting the CycleGAN model to our purpose. CycleGAN is a framework designed to translate between domains with unpaired input-output examples. The architecture assumes some underlying relationship between domains and tries to learn it. Based on a set of images in domain $X$ and a different set in domain $Y$, the algorithm jointly learns a mapping $G: X \rightarrow Y$ and a mapping  $F: Y \rightarrow X$, such that the output $\hat{y} = G(x)$ for every $x \in X$ is indistinguishable from images $y \in Y$, and $\hat{x} = G(y)$ for every $y \in Y$ is indistinguishable from images $x \in X$. Given a mapping $G : X \rightarrow Y$ and another mapping $F : Y \rightarrow X$, then $G$ and $F$ should be one the inverse of the other, and both mappings should be bijections. This property is achieved by training both the mapping $G$ and $F$ simultaneously, and by adding a cycle-consistency loss that encourages $F(G(x))\approx x$ and $G(F(y))\approx y$. Finally, the cycle-consistency loss is combined with the adversarial losses on domains $X$ and $Y$ \cite{zhu2017unpaired}.


\begin{figure*}[]
\centering	
  
\begin{tabular}{cc}
\centering
  \subfloat[Bass discriminator \label{fig:1}]{\includegraphics[width=1\columnwidth]{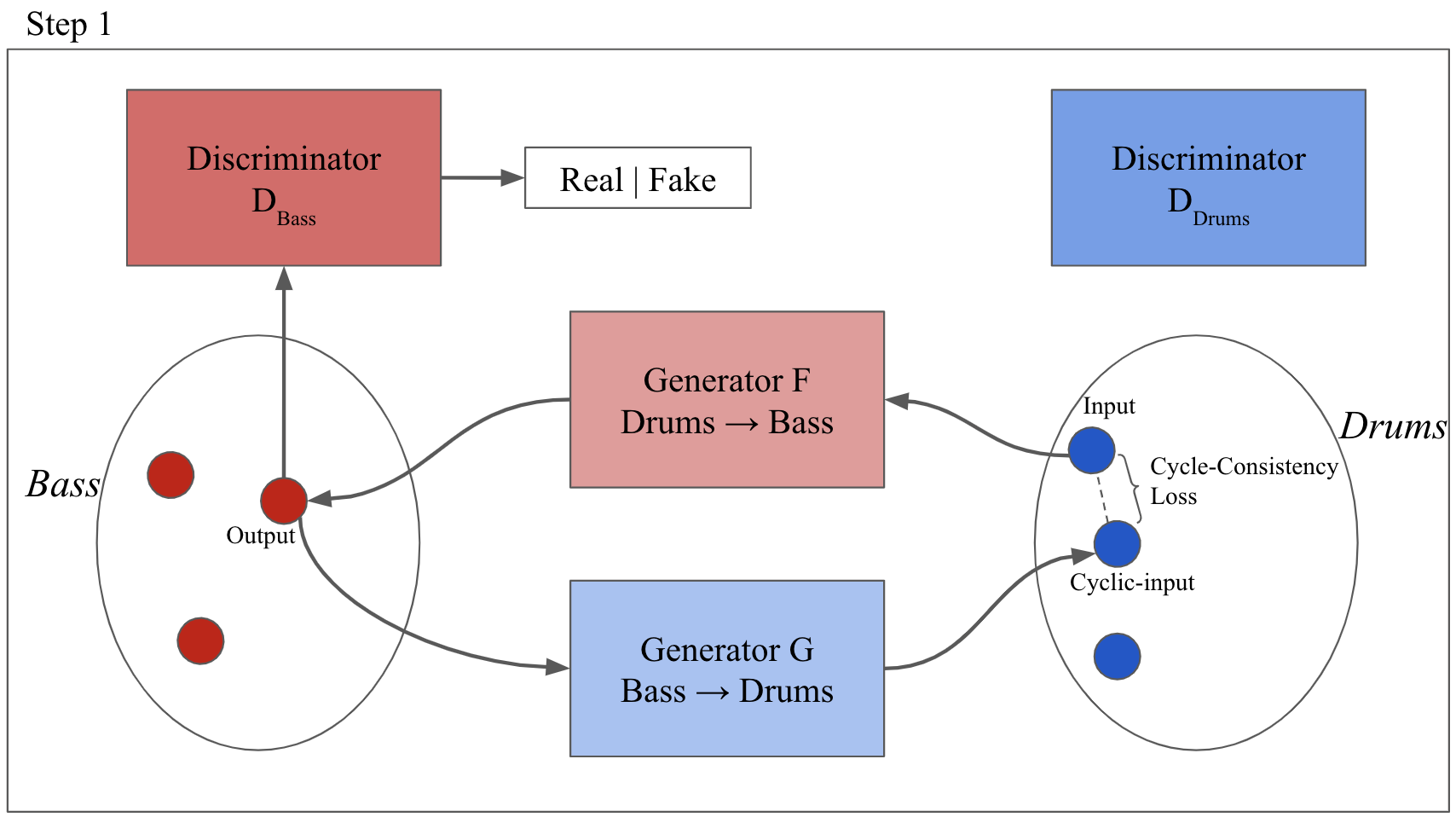}}  & \subfloat[Drum discriminator \label{fig:2}]{\includegraphics[width=1\columnwidth]{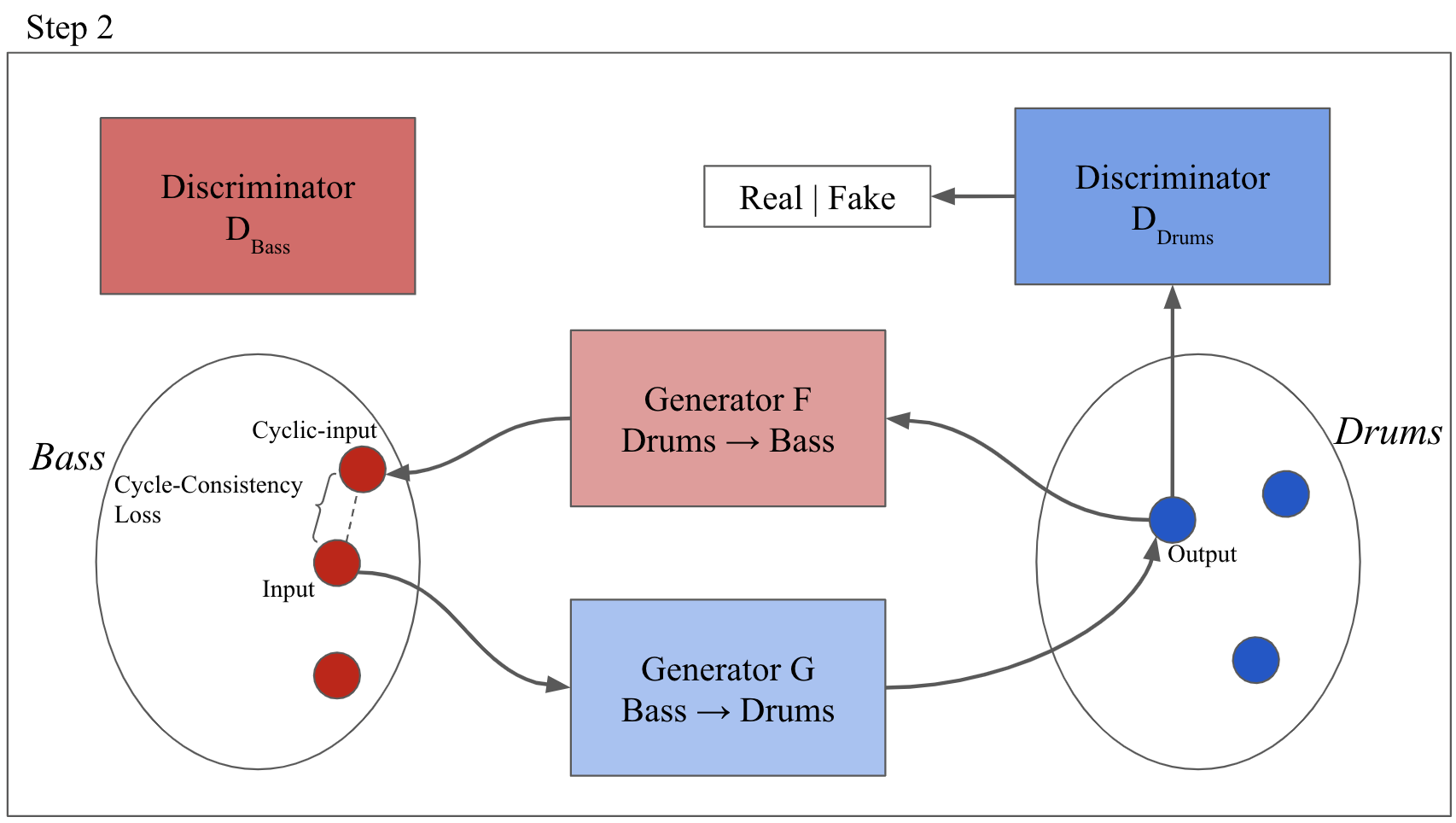}} \\
\end{tabular}
\caption{CycleGAN is a framework designed to translate between domains with unpaired input-output examples. The architecture assumes some underlying relationship between domains and tries to learn it. In our case, based on a set of bass images in domain $X$ and a set of drum images in domain $Y$, the algorithm jointly learns a mapping $G: X \rightarrow Y$ - from bass to drums - and a mapping  $F: Y \rightarrow X$ - from drums to bass, such that the output $\hat{y} = G(x)$ for every $x \in X$ is indistinguishable from images $y \in Y$, and $\hat{x} = G(y)$ for every $y \in Y$ is indistinguishable from images $x \in X$. Given a mapping $G : X \rightarrow Y$ and another mapping $F : Y \rightarrow X$, then $G$ and $F$ should be one the inverse of the other, and both mappings should be bijections. This property is achieved by training both the mapping $G$ and $F$ simultaneously, and by adding a cycle-consistency loss that encourages $F(G(x))\approx x$ and $G(F(y))\approx y$. Finally, the cycle-consistency loss is combined with the adversarial losses on domains $X$ and $Y$ \cite{zhu2017unpaired}.}\label{fig:figs}
\end{figure*}

 


\subsection{Automatic Bass to Drums Arrangement}

CycleDRUMS takes as input a set of $N$ music songs in the waveform domain $X = \{\mathbf{x_{i}}\}_{i=1}^{N}$, where $\mathbf{x_i}$ is a waveform whose number of samples depends on the sampling rate and the audio length. 
Each waveform is then separated by Demucs into different sources. To carry out our experiments, we only used the bass and drum sources. Thus, we ended up having two WAV files for each song, which means a new set of data of the kind: $X_{\text{NEW}} = \{\mathbf{d_{i}}, \mathbf{b_{i}}\}_{i=1}^{N}$, where $\mathbf{b_{i}}, \mathbf{d_{i}}$ represent the bass and drum sources respectively. Each track is then converted into its mel-spectrogram representation.

Since the CycleGAN model takes $256\times256$ images as input, each mel-spectrogram is chunked into smaller pieces with an overlapping window of $50$ time frames, obtaining multiple samples from each song (each equivalent to 5 seconds of music); finally, in order to obtain one channel images from the original spectrograms, we performed a discretization step in the range $[0-255]$. In the final stage of our pipeline, we fed CycleGAN architecture with the obtained dataset. Even though the discretization step introduces some distortion -- original spectrogram values are floats -- the impact on the audio quality is negligible. 

At training time, as the model takes into account two domains $X$ and $Y$, we fed the model with drum and bass lines in order to create credible drums given a bass line.
As previously anticipated, this task is a relevant first step towards fully automated music arrangement. In the future, for instance, this same approach could be applied to more complex signals, such as voice, guitar or piano. Nonetheless, we decided to start from drums and bass because they are usually the first instruments to be recorded when producing a song, and their signals are rather simple compared to more nuanced and harmonic-rich instruments. 

\section{Experiments}

\subsection{Dataset}

For the quality of the generated music samples, it is important to carefully pick the dataset. To train and test our model we decided to use the 
\underline{\textbf{\href{https://github.com/mdeff/fma}{Free Music Archive}}} (FMA), and the 
\underline{\textbf{\href{https://sigsep.github.io/datasets/musdb.html}{musdb18}}} dataset \cite{musdb18} that were both released in 2017. The Free Music Archive (FMA) is the largest publicly available dataset suitable for music information retrieval tasks \cite{fma_challenge}. In its full form it provides 917 GB and 343 days of Creative Commons-licensed audio from 106,574 tracks, 16,341 artists and 14,854 albums, arranged in a hierarchical taxonomy of 161 unbalanced genres. Songs come with full-length and high-quality audio, pre-computed features, together with track- and user-level metadata, tags, and free-form text such as biographies. Given the size of FMA, we chose to select only untrimmed songs tagged as either pop, soul-RnB, or indie-rock, for a total of approximately 10,000 songs ($\approx700$ hours of audio). It is possible to read the full list of songs at \underline{\textbf{\href{https://freemusicarchive.org/home}{FMA website}}}, selecting the genres. We discarded all songs that were recorded live by filtering out all albums that contained the word ``live'' in the title. Finally, in order to better validate and fine-tune our model we decided to also use the full musdb18 dataset. This rather small dataset is made up of 100 tracks taken from the DSD100 dataset, 46 tracks from the MedleyDB, 2 tracks kindly provided by Native Instruments, and 2 tracks from the Canadian rock band The Easton Ellises. It represents a unique and precious source of songs delivered in multi-track fashion. Each song comes as 5 audio files -- vocals, bass, drums, others, and full song -- perfectly separated at the master level. We used the 100 tracks taken from the DSD100 dataset to fine-tune the model ($\approx 6.5$ hours), and the remaining 50 songs to test it ($\approx3.5$ hours). We remark that DEMUCS introduces artifacts in the separated sources output. For this reason, our training strategy is to pre-train the architecture with the artificially source separated FMA dataset, and then fine-tune it with musdb18. Intuitively, the former, which is much larger, helps the model to create a good representation of the musical signal; the latter, which is of higher quality, 
reduces the bias caused by the underlying noise, and favours the automatic generation of a base relying on the (clean) input given only. To conclude, since mel-spectrograms are trimmed in $256\times256$ overlapping windows, we ended up with $600.000$ train samples, and $14.000$ test samples.

\subsection{Training of the CycleGAN model}

We trained our model on 2 Tesla V100 SXM2 GPUs with 32 GB of RAM for 12 epochs (FMA dataset), and fine-tuned it for 20 more epochs (musdb18 dataset). As a final step, the mel-spectrograms obtained were converted to the waveform domain, in order to evaluate the produced music. As to the CycleGAN model used for training, we relied on the default network available at \underline{\textbf{\href{https://github.com/junyanz/pytorch-CycleGAN-and-pix2pix}{this link}}}. 
As a result, the model uses a \textbf{resnet\_9blocks} ResNet generator and a basic 70x70 PatchGAN as a discriminator. The Adam optimizer \cite{kingma2014adam}
was chosen both for the generators and the discriminators, with betas $(0.5, 0.999)$ and learning rate equal to $0.0002$. The batch size was set to 1. The $\lambda$ weights for cycle losses were both equal to 10.

\subsection{Experimental setting}

Even though researchers proposed some effective metrics to predict how popular a song will become \cite{8327835}, there is an intrinsic difficulty in objectively evaluating artistic artifacts such as music. As a human construct, there are no objective, universal criteria for appreciating music.  Nevertheless, in order to establish some forms of benchmark and allow comparisons among different approaches, many generative approaches to raw audio, such as Jukebox \cite{dhariwal2020jukebox}, or Universal Music Translation Network \cite{mor2018universal}, try to overcome this obstacle by having the results manually tagged by human experts. Although this rating may be the best in terms of quality, the result is still somehow subjective, thus different people may end up giving different or biased ratings based on their personal taste. Moreover, the 
cost and time required to manually annotate the dataset could become prohibitive even for relatively few samples (over 1,000). In light of the limits linked to this human-based approach, we propose a new metric that 
correlates well with human judgment. This could represent a first benchmark for the tasks at hand. The 
scores remain somehow subjective, 
as they mirror the evaluators' criteria and grades, but they are obtained based on a fully automatic and standardised approach. 

\subsection{Metrics}
\label{sec:evaluation}

If we consider as a general objective for a system the capacity to assist composers and musicians, rather than to autonomously generate music, we should also consider as an evaluation criteria the satisfaction of the composer, rather than the satisfaction of the auditors \cite{briot2020deep}.

\begin{table}[]
\begin{center}
\caption{Pearson's correlation matrix for all 4 annotators}
\label{tab:IAA}
\begin{tabular}{c|cccc}
           & \textbf{Guitarist} & \textbf{Drummer} & \textbf{Prod. 1} & \textbf{Prod. 2} \\\hline 
\textbf{Guitarist}  & na        & 0.82       & 0.75         & 0.77            \\ 
\textbf{Drummer}    & 0.82          & na      & 0.76            & 0.79            \\ 
\textbf{Prod. 1} & 0.75           & 0.76         & na         & 0.85            \\ 
\textbf{Prod. 2} & 0.77           & 0.79         & 0.85            & na         \\ 
\end{tabular}
\end{center}
\end{table}

However, as previously stated, an exclusive human evaluation may be unsustainable in terms of cost and time required. Thus we carried out the following quantitative assessment of our model. We first produced 400 
samples -- from as many different songs and authors -- of artificial drums starting from bass lines that were 
part of the test set. We then asked a professional guitarist who has been playing in a pop-rock band for more than 10 years, a professional drummer from the same band, and two pop and indie-rock music producers with more than 4 years of experience to manually annotate these samples, capturing the following musical dimensions: sound quality, contamination, credibility, and whether the generated drums followed the beat. More precisely, for each sample, we asked them to rate from 0 to 9 the following aspects: (i) \textit{Sound Quality}: a rating from 0 to 9 of the naturalness and absence of artifacts or noise, (ii) \textit{Contamination}: a rating from 0 to 9 of the contamination by other sources, (iii) \textit{Credibility}: a rating from 0 to 9 of the credibility of the sample, (iv) \textit{Time}: a rating from 0 to 9 of whether the produced drums follow the beat of the bass line. The choice fell on these four aspects after we asked the evaluators to list and describe the most relevant dimensions in the perceived quality of drums. The correlation matrix for all 4 annotators is shown in Table \ref{tab:IAA}.

Ideally, we want to produce some quantitative measure whose outputs -- when applied to generated samples -- 
correlate well (i.e. predict) expert average grades. To achieve this goal, we trained a logistic regression model with features obtained through a comparison between the original drums and the artificial drums. Here are the details on how we obtained suitable features.

\textbf{STOI-like features.} We created a procedure -- inspired by the STOI \cite{andersen2017non} -- whose output vector somehow measures the mel-frequency bins correlation throughout time between the original sample and the fake one. The obtained vector can then be used to feed a multi regression model whose independent variable is the human score attributed to that sample.  Here is the formalisation: $$
HumanScore = \sum_i^{256} a_i \biggl[ \sum_t^{256}(x_i^{(t)}-\bar x^{(t)})(y_i^{(t)}-\bar y^{(t)}) \biggr]
$$

To simplify, to each pair of samples (original and generated one) a $256$ element long vector is associated as follows: 

$$\mathcal{S}(\mathcal{X},\mathcal{Y}, l)^{(i)} = \sum_t^{256}(x_i^{(t)}-\bar x^{(t)})(y_i^{(t)}-\bar y^{(t)})
$$

Where: (i.) $\mathcal{X}$ and $\mathcal{Y}$ are, respectively, the mel-spectrogram matrices of original and generated samples; (ii.) $a_i$ is the $i$-th coefficient for the linear regression; (iii.) $x_i^{(t)}$ and $y_i^{(t)}$ the $i$-th element of the $t$-th column of matrices $\mathcal{X}$ and $\mathcal{Y}$, respectively; (iv.) $\bar x^{(t)}$ and $\bar y^{(t)}$ are the means along the $t$-th column of matrices $\mathcal{X}$ and $\mathcal{Y}$, respectively. Each feature $i$ of the regression model is a sort of Pearson correlation coefficient between row $i$ of $\mathcal{X}$ and row $i$ of $\mathcal{Y}$ throughout time.

\textbf{FID-based features.} In the context of GANs result evaluation, the Fréchet Inception distance (FID) is supposed to improve on the Inception Score by actually comparing the statistics of generated samples to real samples 
\cite{heusel2017gans}. In other words, FID measures the probabilistic distance between two multivariate Gaussians, where $X_r = N(\mu_r,\Sigma_r)$ and $X_g = N(\mu_g,\Sigma_g)$ are the 2048-dimensional activations of the Inception-v3 pool3 layer -- for real and generated samples respectively -- modeled as normal distributions. The similarity between the two distributions is measured as follow:

$$
FID=||\mu_r - \mu_g||^2+Tr(\Sigma_r+\Sigma_g - 2(\Sigma_r\Sigma_g)^{1/2})
$$ 

Nevertheless, since we want to assign a score to each sample, we just estimated the  $X_r = N(\mu_r,\Sigma_r)$ parameters -- using different activation layers of the Inception pre-trained network -- and then we calculated the probability density associated to each fake sample. Finally, we added these scores to the regression model predictors.

\subsection{Baseline}

\begin{figure*}[!ht]
\begin{center}
\includegraphics[width=0.7\textwidth]{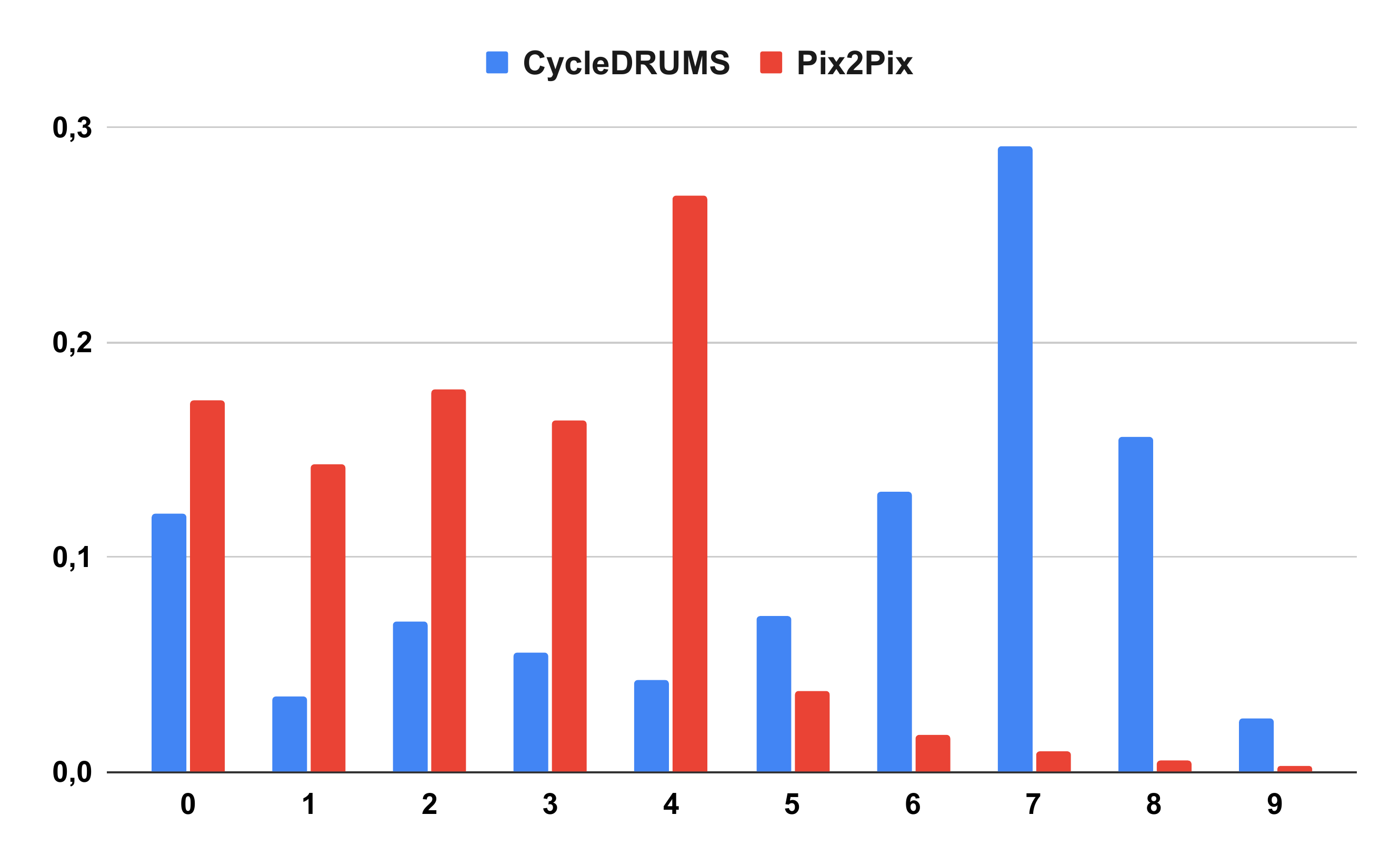}
\end{center}
\caption{\centering   The distribution of grades for the 400 test drums for both CycleGAN and Pix2Pix (baseline) -- averaged among all four independent evaluators and over all four dimensions. We rounded the results to the closest integer to make the plot more readable. The higher the grade, the better the sample will sound. Samples with grade 0-3 are generally silent or very noisy. In samples graded 4-5 few sounds start to emerge, but they are usually not very pleasant to listen to, nor coherent. Grades 6-7 identify drums that sound good, that are coherent, but that are not continuous: they tend to follow the bass line too closely. Finally, samples graded 8 and 9 are almost indistinguishable from real drums, both in terms of sound and timing.}
\label{fig:grade_dist}
\end{figure*}

Since, to the best of our knowledge, we are the first to tackle the drum arrangement task in the audio domain, and to treat it as an image-to-image translation problem, we 
lack of a suitable baseline. In the end, instead of forcing a pre-existing method to work in our specific scenario, we decided to replicate our experiments using the Pix2Pix architecture \cite{isola2017image}, another image-to-image translation network. Unlike CycleGAN, Pix2Pix learns to translate between domains when fed with paired input-output examples. At training time, we relied on the default network available \underline{\textbf{\href{https://github.com/junyanz/pytorch-CycleGAN-and-pix2pix}{here}}}, we run it on 2 Tesla V100 SXM2 GPUs with 32 GB of RAM for 50 epochs (FMA dataset), and we fine-tuned it for 30 more epochs (musdb18 dataset). 

Finally, after training was completed, we produced 400 drum samples from the same bass lines used for generating the test drums that the evaluators graded. We then asked the same four evaluators to grade the new drum samples according to the principles presented in section \ref{sec:evaluation}. 

\subsection{Experimental Results}

Figure \ref{fig:grade_dist} shows the distribution of grades for the 400 test drums for both CycleGAN and Pix2Pix -- averaged among all four independent evaluators and over all four dimensions. We rounded the results to the closest integer to make the plot more readable. The higher the grade, the better the sample will sound. Additionally, to fully understand what to expect from samples graded similarly, we discussed the model results with the evaluators. We collectively listened to a random set of samples and it turned out that all four raters followed similar principles in assigning the grades. Samples with grade 0-3 are generally silent or very noisy. In samples graded 4-5 few sounds start to emerge, but they are usually not very pleasant to listen to, nor coherent. Grades 6-7 identify drums that sound good, that are coherent, but that are not continuous: they tend to follow the bass line too closely. Finally, samples graded 8 and 9 are almost indistinguishable from real drums, both in terms of sound and timing. In labeling non graded samples, we therefore trained a multi-logistic regression model with both the STOI-like and the FID-based features to predict what of these four buckets the graders would assign the sample to. The model accuracy on test set was 87\% for CycleDRUMS and 93\% for Pix2Pix.

\begin{figure*}[!ht]
\begin{center}
\includegraphics[width=0.7\textwidth]{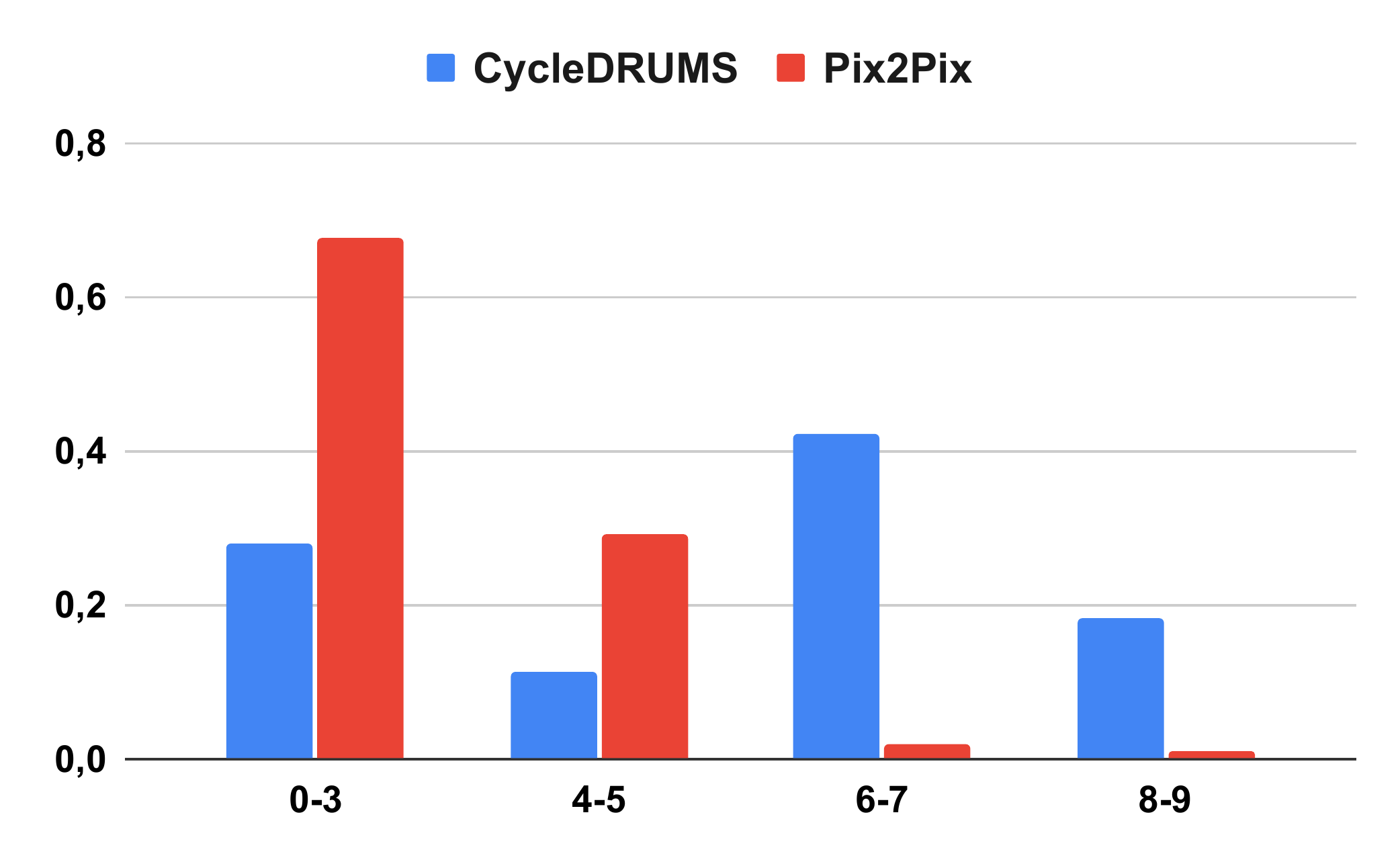}
\end{center}
\caption{\centering In labeling non graded samples, we therefore trained a multi-logistic regression model with both the STOI-like and the FID-based features to predict what of the four grade buckets the graders would assign the sample to. The model accuracy on test set was 87\% for CycleDRUMS and 93\% for Pix2Pix. Given this pretty good result, we could then use this trained logistic model to label 14000 different 5s fake drum clips, produced from as many real bass lines using both CycleGAN and Pix2Pix (baseline). The histogram above shows the distribution of predicted class for these samples.} 
\label{fig:classes}
\end{figure*}

Given this pretty good result, we could then use this trained logistic model to label 14000 different 5s fake drum clips, produced from as many real bass lines using both CycleGAN and Pix2Pix. Figure \ref{fig:classes} shows the distribution of predicted class for these samples. At \underline{\textbf{\href{https://soundcloud.com/user-639025674/sets/bass2drums/s-jjccrgdXXOi}{this website}}} a private Sound Cloud playlist of some of the most interesting results is available, while at \underline{\textbf{\href{https://soundcloud.com/user-639025674/sets/bass2drums_baseline/s-rKqV2S2MFHw}{this one}}} we uploaded some samples obtained with the Pix2Pix baseline architecture. 

Finally, with respect to the computational resources and time required to generate new arrangements, our approach shows several advantages, compared to auto-regressive models \cite{dhariwal2020jukebox}. Since the output prediction can be fully parallelized, the inference time amounts to a forward pass and a Mel-spectrogram-waveform inverse conversion, whose duration depends on the input length, but it never exceeds few minutes. Indeed, it is worth noting that, at inference time, arbitrary long inputs can be processed and arranged.

\section{Conclusions and Future Work}

In this work, we presented a novel approach to automatically produce drums starting from a bass line. We applied CycleGAN to real bass lines, treated as gray-scale images (mel-spectrograms), obtaining good ratings, especially if compared to another image-to-image translation approach (Pix2pix). Given the novelty of the problem, we proposed a reasonable procedure to properly evaluate our model outputs. Notwithstanding the promising results, some critical issues need to be addressed before a more compelling architecture can be developed. First and foremost, a larger and cleaner dataset of source separated songs should be created. In fact, manually separated tracks always contain a big deal of noise. Moreover, the model architecture should be further improved to focus on longer dependencies and to take into account the actual degradation of high frequencies. For example, our pipeline could be extended to include some recent work on quality-aware image-to-image translation networks \cite{8673660}, and spatial attention generative adversarial networks \cite{9007501}. Finally, a certain degree of interaction and randomness should be inserted to make the model less deterministic and to give creators some control over the sample generation.  
Our contribution is nonetheless a first step toward more realistic and useful automatic music arrangement systems and we believe that further significant steps could be made to reach the final goal of human-level automatic music arrangement production. 
Moreover, this task moves towards the direction of automatic music arrangement (the same methodology could possibly be extended, in future, to more complex domains, such as voice or guitar or the whole song). 
Already now software like Melodyne \cite{neubacker2011sound,senior2009celemony}
delivers producers a powerful user interface to directly 
modify and adjust a spectrogram-based representation of audio signals to correct, perfect, reshape and restructure vocals, samples and recordings of all kinds. 
It is not unlikely that in the future artists and composers will start creating their music almost like they were drawing.


%





\ifCLASSOPTIONcaptionsoff
  \newpage
\fi



\bibliographystyle{IEEEtran}
\bibliography{journal.bib}
\end{document}